\documentclass[a4paper,11pt]{article}
\usepackage{pos}
\usepackage{psfrag,slashed,cancel,array,graphicx,todonotes,hyperref}
\usepackage{mathrsfs}
\usepackage{slashed}
\usepackage{hhline}
\usepackage{multirow}   
\newcommand{\nn}{\nonumber}
\newcommand{\as}{\alpha_s}
\newcommand{\cT}{\mathcal{T}}
\newcommand{\eps}{\epsilon}
\newcommand{\LQCD}{\Lambda_\mathrm{QCD}}

\title{NNLO DGLAP splitting functions from  collinear matching of TMDs}

\author*{Yu Jiao Zhu}

\affiliation{Max-Planck-Institut f\"{u}r Physik, Werner-Heisenberg-Institut, \\
Boltzmannstr. 8, 85748 Garching, Germany}

\emailAdd{yzhu@mpp.mpg.de}

\abstract{
We report a complete computation of next-to-next-to-leading order (NNLO) helicity and transversity Dokshitzer-Gribov-Lipatov-Altarelli-Parisi (DGLAP) splitting functions, in both space-like and time-like kinematics.
These results are obtained from  the next-to-next-to-next-to-leading order (N$^3$LO) twist-2 matching of polarized transverse-momentum-dependent (TMD) parton distribution and fragmentation functions, including helicity, quark transversity, and linearly polarized gluons.
We compare our results with existing calculations in the literature and discuss both agreements and discrepancies.
Our results provide all perturbative ingredients required for the  computation of N$^3$LO differential cross sections below the resolution scale $q_{T\mathrm{cut}}$ in transverse-momentum subtraction
and  enable next-to-next-to-next-to-next-to-leading logarithmic (N$^4$LL) resummation of 
 $q_T$ observables in the Sudakov region.
We further determine the  small-$x$ structure of the polarized matching coefficients through N$^3$LO. 
These fixed-order results furnish the data for future small-$x$ resummation in polarized TMD factorization, 
where high-energy logarithms and Sudakov logarithms become simultaneously relevant. 
Establishing a consistent joint treatment of polarized small-$x$ evolution and transverse-momentum resummation remains an important open direction toward uniform precision in spin-dependent phenomenology.
Our results  provide essential theoretical input for precision spin physics at the forthcoming Electron-Ion Collider.
}

\FullConference{17th International Symposium on Radiative Corrections: Applications of Quantum Field Theory to Phenomenology (RADCOR2025)\\
5-10 October 2025\\
Puri, India\\}

\begin{document}
\maketitle

\section{Introduction}
\label{sec:introduction}
Transverse-Momentum-Dependent (TMD) Parton Distribution Functions (PDFs)
parameterize the confined motion of parton constituents inside a nucleon.
TMD Fragmentation Functions (FFs) describe the hadronization process of partons into color-singlet hadrons.
Beyond their role in transverse-momentum spectra,
TMDs encode multidimensional information about the nucleon’s internal flavor and spin structure.
In particular, small-$x$ TMDs probe the Regge asymptotics of QCD governed by
Balitsky-Fadin-Kuraev-Lipatov (BFKL) evolution~\cite{Balitsky:1978ic,Kuraev:1977fs} and its generalizations.
TMDs are therefore key objects in the next QCD precision frontier
and play a central role in understanding how quarks and gluons form hadronic matter~\cite{Accardi:2012qut}.

A longstanding objective in QCD is to uncover the fundamental mechanism of color confinement.
At the same time, QCD is characterized by the converse property of asymptotic freedom~\cite{Gross:1973id,Politzer:1973fx}.
The hierarchy between short- and long-distance scales allows these two regimes to be systematically disentangled within effective field theory.
Schematically, a high-energy cross section admits the factorized structure
\begin{align}
\sigma (Q,\Lambda_{\mathrm{QCD}})= 
\mathcal{C}(Q,\mu)\otimes \langle \mathcal{O} \rangle (\mu,\Lambda_{\mathrm{QCD}}) 
+ \mathcal{O}\!\left(\frac{\Lambda^2_{\mathrm{QCD}}}{Q^2}\right),
\end{align}
where $\mu$ denotes the factorization scale.
Within the Collins-Soper-Sterman (CSS) framework~\cite{Collins:1988ig},
such factorization has been established for Drell-Yan and single inclusive hadron production~\cite{Bodwin:1984hc,Collins:1984kg,Collins:1989gx,Nayak:2005rt}.

Factorization underlies all precision predictions in QCD.
To the UV, the renormalization of $\alpha_s$ is independent of the external momenta flowing into Green functions,
so that UV physics is completely factorized from collider measurements performed on asymptotic final states.
To the IR, long-distance physics is parametrized while short-distance dynamics are computed
from Feynman amplitudes, again thanks to asymptotic freedom.
UV and IR cutoff dependence drop out in the \emph{relations} among physical quantities.

In the TMD context, factorization bridges partonic dynamics to experimentally measurable observables.
The intrinsic motion of confined partons cannot be accessed directly;
it manifests through their correlations with final-state leptons, gauge bosons, and hadrons, owing to the unification of Standard Model gauge group ${\text{SU}}(3)\times {\text{SU}}(2)\times {\text{U}}(1)_{\text{Y}}~({\text{SU}}(2)\times {\text{U}}(1)_{\text{Y}}\to\text{U}(1)_{\text{EM}})$. Indeed, TMDs are indispensable non-perturbative input for a variety of benchmark observables in the Standard Model, including Drell-Yan process~\cite{ Dokshitzer:1978yd, Parisi:1979se, Collins:1984kg, Arnold:1990yk, Ladinsky:1993zn, Bozzi:2010xn, Becher:2010tm, Becher:2011xn, Bertone:2019nxa}, semi-inclusive deep-inelastic scattering (SIDIS)~\cite{Ji:2004wu, Ji:2004xq, Liu:2018trl, Fang:2024auf, Gao:2022bzi, Ebert:2021jhy,Fang:2024auf}, electron-positron annihilation to hadrons and jets~\cite{Collins:1981uk,Collins:1981va,Neill:2016vbi,Gutierrez-Reyes:2018qez,Gutierrez-Reyes:2019vbx,Gutierrez-Reyes:2019msa}, Higgs boson production~\cite{Berger:2002ut,Bozzi:2005wk,Gao:2005iu,Echevarria:2015uaa,Neill:2015roa,Bizon:2017rah,Chen:2018pzu,Bizon:2018foh}, top quark pair production~\cite{Zhu:2012ts,Li:2013mia,Catani:2014qha,Catani:2018mei,Dai:2026hso}, $J/\psi$ production~\cite{Echevarria:2024idp, Copeland:2023wbu, Kishore:2018ugo, Maxia:2024cjh,Banu:2024ywv,Kishore:2024bdd,Chakrabarti:2022rjr,Boer:2023zit,Bor:2022fga,Bacchetta:2018ivt,Boer:2016bfj}, as well as Energy-Energy (EEC) or Charge-Charge Correlator~\cite{Moult:2018jzp,Gao:2019ojf,Gao:2024wcg,Kang:2024otf,Kang:2023big,Liu:2024kqt,Monni:2025zyv} at lepton and hadron colliders.

Polarized TMDs are particularly important for addressing the long-standing proton spin problem.
The proton spin crisis revealed that quark helicity accounts for only a fraction of the nucleon spin,
raising fundamental questions about the roles of gluon helicity and orbital angular momentum.
Helicity TMDs probe longitudinal spin-momentum correlations,
quark transversity distributions encode transverse spin structure,
and linearly polarized gluon TMDs access interference between gluon helicity states.
Together, these observables provide a multidimensional tomography of spin and motion inside the nucleon.
Achieving quantitative precision in spin physics requires perturbative control over their matching and evolution.

Recent years have witnessed major progress in computing higher-order radiative corrections for the perturbative part of unpolarized TMDs~\cite{Luo:2019szz,Luo:2020epw,Ebert:2020qef,Ebert:2020yqt}.
The polarized sector, however, has lagged behind in perturbative accuracy.
Here we complete the N$^3$LO twist-2 matching program for the leading polarized TMDs~\cite{Zhu:2025gts,Zhu:2025ixc,Zhu:2025brn}:
helicity, quark transversity, and linearly polarized gluon distributions and fragmentation functions.
The calculation is performed within Soft-Collinear Effective Theory~\cite{Bauer:2000ew,Bauer:2000yr,Bauer:2001ct,Bauer:2001yt} with an exponential rapidity regulator~\cite{Li:2016axz},
allowing for a systematic treatment of rapidity divergences.  These results enable the computation of N$^3$LO differential cross sections below the resolution scale $q_{T\mathrm{cut}}$ in transverse-momentum subtraction~\cite{Catani:2007vq,Catani:2009sm,Catani:2010en,Catani:2011qz} and N$^4$LL resummation
of  $q_T$ observables in the Sudakov region.

We also extract the complete set of NNLO  DGLAP splitting functions
in both space-like and time-like kinematics, and the  small-$x$ behavior of the matching coefficients (and DGLAP splitting functions) through N$^3$LO (and NNLO).
These fixed-order results provide the necessary perturbative input for future small-$x$ resummation
in kinematic regimes where high-energy logarithms and Sudakov logarithms can coexist.
Developing a consistent framework that simultaneously resums small-$x$ and transverse-momentum logarithms
remains an important step toward achieving uniform precision in spin-dependent observables.

In this talk, I present an overview of recent progress in the collinear matching of TMDs and the associated polarized DGLAP splitting functions.
\section{Operator definitions}
The TMD  parton distribution  functions can be defined in terms of SCET
gauge invariant collinear fields
\begin{align}
  \label{eq:PDFdef}
   \mathcal{B}^{0 ij}_{q/N}(x,b_\perp) = &
\int \frac{db^-}{2\pi} \, e^{-i x b^- P^{+}} 
 \langle N(P);S_{\parallel} | \bar{\chi}^i_n(0,b^-,b_\perp) \chi^j_n(0) | N(P);S_{\parallel} \rangle \, ,
\nonumber\\ 
  {\cal B}_{g/N}^{ 0\mu \nu}(x,b_\perp) =&
 \,  x P_+ \int \frac{d b^-}{2 \pi} e^{- i x b^- P^+} \langle N (P);S_{\parallel} | {\cal A}_{n \perp}^{a,\mu} (0, b^-, b_\perp) {\cal A}_{n \perp}^{a,\nu}(0) | N(P);S_{\parallel} \rangle \,.
\end{align}
Here $N(P)$ is a target with momentum $P^\mu = (\bar{n} \cdot P) n^\mu/2 = P^+ n^\mu/2$,  where
 \begin{align}
n^\mu = (1, 0, 0, 1)\,,\quad \bar{n}^\mu = (1, 0, 0, -1)\,.
\end{align}
 $\chi_n = W_n^\dagger \xi_n$ is the gauge invariant collinear quark field in SCET, 
constructed from collinear quark field $\xi_n$ and path-ordered collinear Wilson line 
\begin{align}
W_n(x) = {\cal P} \exp \left(i g \int_{-\infty}^0 ds\, \bar{n} \cdot A_n (x + \bar{n} s) \right)\,.
\end{align}
$\mathcal{A}_{n\perp}^{a,\mu}$ is the gauge invariant collinear gluon field with color index $a$ and Lorentz index $\mu$, 
constructed from the collinear gluon field and path-ordered collinear Wilson line in the  adjoint representation.

In this work, we  consider longitudinally polarized target having chirality $S_{\parallel}=\pm$, so that only a part of the complete spin decomposition is needed, we have, for the quark TMD PDFs
\begin{align}
 {\cal B}_{q/N}^{0 i j}(x,b_\perp) =\frac{1}{2}\left(
  \frac{\slashed n^{ij}}{2} \mathcal{B}^{0q}_{1}(x, b_T)+S_{\parallel} \frac{(\slashed n\gamma_5)^{ij}}{2} \Delta \mathcal{B}^{0q}_{1}(x, b_T)
 +\dots
 \right)\,,
\end{align}
where
\begin{align}
\mathcal{B}^{0 q}_{1}(x, b_T)=
\mathrm{Tr}\left [\frac{\slashed {\bar n} }{2} {\cal B}^0_{q/N}(x,b_\perp)\right]\,,\quad
 \Delta \mathcal{B}^{0q}_{1}(x, b_T)=
 \mathrm{Tr}\left [\frac{\slashed {\bar n } }{2}\gamma_5 {\cal B}^0_{q/N}(x,b_\perp,S_{\parallel}=+)\right]\,.
\end{align}
And  for the gluon TMD PDFs
\begin{align}
 {\cal B}_{g/N}^{ 0\mu \nu}(x,b_\perp) =\frac{1}{2}\left(
  - g_\perp^{\mu\nu}
\mathcal{B}^{0g}_{1}(x, b_T) - \left(\frac{g_\perp^{\mu\nu}}{2} + \frac{b_\perp^\mu b_\perp^\nu}{b_T^2} \right) 2
\mathcal{H}^{0\perp g}_{1}(x, b_T)+
S_{\parallel}(-i\epsilon_\perp^{\mu\nu}) \Delta \mathcal{B}^{0g}_{1}(x, b_T) 
+\dots\right)\,,
\end{align}
where 
\begin{align}
\mathcal{B}^{0g}_{1}(x, b_T) =&-g_{\perp{\mu\nu}}{\cal B}_{g/N}^{0\mu \nu}(x,b_\perp)\,,\quad
\mathcal{H}^{0\perp g}_{1}(x, b_T)=
 -\left(  g_{\perp\mu\nu} +2\frac{b_{\perp\mu} b_{\perp\nu}}{b_T^2} \right) {\cal B}_{g/N}^{ 0\mu \nu}(x,b_\perp)\,,
 \nn\\
 \Delta \mathcal{B}^{0g}_{1}(x, b_T) =&\,
 i\epsilon_{\perp{\mu\nu}}{\cal B}_{g/N}^{0\mu \nu}(x,b_\perp,S_{\parallel}=+)\,,
\end{align}
and
\begin{align}
g_\perp^{\mu\nu}=g^{\mu\nu}-\frac{n^\mu \bar n^\nu+ n^\nu \bar n^\mu}{n\cdot\bar n}\,,
\quad
\epsilon_\perp^{\mu\nu}=\frac{n_\sigma\bar n_\tau}{n\cdot\bar n}\epsilon^{\mu\nu\sigma\tau}\,.
\end{align}
In the hadron frame, where the detected hadron has zero transverse momentum, an operator definition for gluon TMD FFs can be written down
\begin{align}
  \label{eq:FFdef}
   \mathcal{D}^{0 ij}_{N/q}(z,b_\perp) =&
    \frac{1}{z N_c}
\int_X\int \frac{db^-}{2\pi} \, e^{-i  b^- P^{+}/z} 
 \langle 0|T [\chi^j_n(0,b^-,b_\perp)]| N(P); X\rangle 
 \langle   N(P); X | \bar T[\bar{\chi}^i_n(0)]|0\rangle\,,
 \nonumber\\ 
  {\cal D}_{N/g}^{ 0\mu \nu}(z,b_\perp) =& \frac{P_+}{ z^2(N^2_c-1)}
 \,   \int_X\int\frac{d b^-}{2 \pi} e^{- i  b^- P^+/z} \langle 0|T [ {\cal A}_{n \perp}^{a,\mu} (0, b^-, b_\perp)]| N(P); X\rangle 
 \nn\\
 \times &
 \langle   N(P); X | \bar T[ {\cal A}_{n \perp}^{a,\nu}(0) ]|0\rangle \,.
\end{align}
In practical calculations, in particular for FF renormalization, it's also convenient to define the fragmentation functions in the parton frame,
 where the parton which initiates the fragmentation has zero transverse momentum.
  The parton frame TMDFFs are related to the hadron frame ones by~\cite{Collins:2011zzd,Luo:2019hmp,Luo:2019bmw}
\begin{align}
  \label{eq:partontohadron}
  \mathcal {F}^0_{N/i} (z, b_\perp/z) = z^{2 - 2 \epsilon} \mathcal{D}^0_{N/i} (z, b_\perp)  \,,
\end{align}
where we denote the bare partonic TMD FFs in the parton frame by $\mathcal {F}_{N/i}$. 
Our  results for TMD FFs will be given in the hadron frame, by choosing the argument of the parton frame coefficient to be $b_\perp/z$. 
Similar to the TMD PDFs, we have spin decomposition for TMD FFs as follows
 \begin{align}
\mathcal{D}^{0 ij}_{N/q}(z,b_\perp) =
\frac{ \slashed n^{ij}}{2}\mathcal{D}^{0q}_{1}(z, b_T)
 +\dots
 \,,
\end{align}
where
\begin{align}
\mathcal{D}^{0q}_{1}(z, b_T)=
\frac{1}{2}\mathrm{Tr}\left [\frac{\slashed {\bar n} }{2} \mathcal{D}^{0}_{N/q}(z,b_\perp) \right]\,.
\end{align}
And  for the gluon TMD FFs
\begin{align}
 {\cal D}_{N/g}^{0 \mu \nu}(z,b_\perp) =
  - g_\perp^{\mu\nu}
\mathcal{D}^{0g}_{1}(z, b_T) - \left(\frac{g_\perp^{\mu\nu}}{2} + \frac{b_\perp^\mu b_\perp^\nu}{b_T^2} \right) 2
\mathcal{H}^{0\perp g}_{1}(z, b_T)
+\dots\,,
\end{align}
where
\begin{align}
\mathcal{D}^{0g}_{1}(z, b_T)=&-\frac{1}{2}g_{\perp{\mu\nu}}{\cal D}_{N/g}^{ 0\mu \nu}(z,b_\perp)\,,\quad
\mathcal{D}^{0g}_{1}(z, b_T)=
-\frac{1}{2} \left(  g_{\perp\mu\nu} +2\frac{b_{\perp\mu} b_{\perp\nu}}{b_T^2} \right) {\cal D}_{N/g}^{0 \mu \nu}(z,b_\perp)\,.
\end{align}
The collinear functions defined so far correspond to unsubtracted beam functions. 
The genuine collinear functions are obtained by subtracting the soft contributions from the collinear sector~\cite{Collins:1982wa,Collins:2011zzd,Echevarria:2012js,Ji:2004wu}
\begin{align}
    \mathcal{B}_{i/N}& (x\,,b_\perp\,,E_n, \mu,\nu)
    \equiv
    \lim_{\nu\to\infty}
    Z^i_B(b_\perp\,,E_n, \mu,\nu)\frac{\mathcal{B}^{0}_{i/N} (x\,,b_\perp\,,E_n, \mu,\nu)}{\mathcal{S}^{0}_{n\bar n}(b_\perp\,,\mu,\nu)}\,,
\end{align}
where the superscript $0$ denotes  unsubtracted collinear beam functions, $Z^i_B$ is the UV renormalization factor, $i=q,g$.
Here, $\nu$ is rapidity factorization scale to separate soft and collinear contribution 
~\cite{Collins:2011zzd,Becher:2010tm,Echevarria:2011epo,Echevarria:2012js,
Chiu:2012ir,Li:2016axz,Echevarria:2015usa,Echevarria:2015byo,Echevarria:2016scs,Becher:2011dz,Ebert:2018gsn}.
In this work, we adopt exponential regulator~\cite{Li:2016axz} for both soft~\cite{Li:2016ctv} and collinear sectors~\cite{Luo:2019bmw,Luo:2019hmp,Luo:2019szz,Luo:2020epw}, as 
it explicitly preserves non-abelian exponentiation~\cite{Gatheral:1983cz,Frenkel:1984pz}, 
and the zero-bin soft function within exponential regularization scheme is identical to the TMD soft function.
After zero-bin subtraction, the physical TMDs are obtained by dressing the collinear functions with the cloud of soft gluons
\begin{align}
f_1^i(x,b_\perp,\mu,\zeta)=\,& {\cal B}_{i/N} (x,b_\perp, E_n,\mu,\nu)\sqrt{\mathcal{S}_{n\bar n}(b_\perp\,,\mu,\nu)}\,,
\nn\\ 
g_{1L}^i(x,b_\perp,\mu,\zeta)=\,&\Delta {\cal B}_{i/N} (x,b_\perp, E_n,\mu,\nu)\sqrt{\mathcal{S}_{n\bar n}(b_\perp\,,\mu,\nu)}\,,
\nn\\ 
h^{\perp g}_{1}(x,b_\perp,\mu,\zeta)=\,&\mathcal{H}^{\perp g}_{1}(x,b_\perp, E_n,\mu,\nu)\sqrt{\mathcal{S}_{n\bar n}(b_\perp\,,\mu,\nu)}\,.
\end{align}
The rapidity cut-off dependence cancel between soft and collinear sectors, leaving physical rapidity scales known as CS scales~\cite{Collins:1981uk,Collins:1981uw,Collins:1984kg}.
The rapidity evolution  was pioneered by Collins, Soper, and Sterman (CSS),
and subsequently extended in Refs.~\cite{Collins:2011zzd,Ji:2004wu,deFlorian:2001zd,Catani:2000vq,Catani:2010pd,Aybat:2011zv,Ji:2005nu}.
Later, this formalism was recast in terms of rapidity renormalization group equations (RRGs)~\cite{Chiu:2011qc,Chiu:2012ir} within the framework of SCET~\cite{Becher:2010tm,Becher:2011xn,Becher:2012yn,Echevarria:2011epo,Echevarria:2012js,Echevarria:2014rua,Chiu:2012ir,Chiu:2011qc,Ebert:2019tvc}.
 \section{The collinear matching of TMDs and NNLO DGLAP evolutions}
The collinear matching for unpolarized, helicity, linearly polarized gluon, and quark transversity TMDs is now known through N$^3$LO in Refs.~\cite{Luo:2019szz,Luo:2020epw,Ebert:2020qef,Ebert:2020yqt,Zhu:2025brn,Zhu:2025ixc,Zhu:2025gts}, for both beam and fragmentation functions.
We note minor discrepancies compared to the work of Ref.~\cite{Echevarria:2015usa,Gutierrez-Reyes:2018iod}.
The rapidity evolution of the TMDs has reached N${}^4$LL accuracy~\cite{Li:2016ctv,Moult:2022xzt,Duhr:2022yyp}.
These results have recently led to a notable advance toward NNLO $+$ N${}^4$LL QCD predictions for hadron production in DIS at finite transverse momentum~\cite{Gao:2026tnd, Dong:2026eas}.
Below we update the table, adapted from Ref.~\cite{Moos:2020wvd,Boussarie:2023izj}, to reflect the current status of collinear matching for quark  and gluon TMD PDFs.
For brevity, the renormalization scale $\mu$ and the Collins-Soper scale $\zeta$ are suppressed.
\begin{table}
 \centering
 \begin{tabular}{|l|c|c|c|r|c|}
 \hline
 \multirow{2}{*}{Name} & \multirow{2}{*}{Function} & Twist-2  & Twist-3  & \multicolumn{1}{c|}{Known}   & \multirow{2}{*}{Refs.} \\
 & & matching & matching & \multicolumn{1}{c|}{order} &
 \\ \hhline{|=|=|=|=|=|=|}
 unpolarized   & $\tilde f_1(x,b_T)$          & $f_1(x)$ & --       & N$^3$LO $(\as^3)$ &  \cite{Luo:2019szz, Luo:2020epw, Ebert:2020yqt}
 \\ \hline
 helicity      & $\tilde g_{1}(x,b_T)$       & $g_1(x)$ & $\cT_g(x)$ & N$^3$LO $(\as^3)$     & \cite{Zhu:2025gts}
 \\ \hline
 worm-gear $T$ & $\tilde g_{1T}^\perp(x,b_T)$ & $g_1(x)$ & $\cT_g(x)$ & LO $(\as^0)$      & \cite{Kanazawa:2015ajw, Scimemi:2018mmi}
 \\ \hline
 Sivers        & $\tilde f_{1T}^\perp(x,b_T)$ & --       & $T(-x,0,x)$ & NLO $(\as^1)$     & \cite{Boer:2003cm,Ji:2006ub, Ji:2006vf, Koike:2007dg}
\\ & & & & &
 \cite{Kang:2011mr, Sun:2013hua, Dai:2014ala, Scimemi:2019gge,Rein:2022odl}
 \\ \hhline{|=|=|=|=|=|=|}
 transversity  & $\tilde h_1(x,b_T)$          & $h_1(x)$ & $\cT_h(x)$ & N$^3$LO $(\as^3)$ & \cite{Zhu:2025brn}
 \\ \hline
 worm-gear $L$ & $\tilde h_{1L}^\perp(x,b_T)$ & $h_1(x)$ & $\cT_h(x)$ & NLO $(\as^1)$   & \cite{Rein:2022odl}
 \\ \hline
 Boer-Mulders  & $\tilde h_1^\perp(x,b_T)$    & --       & $\delta T_\eps(-x,0,x)$ & NLO $(\as^0)$ & \cite{Scimemi:2018mmi,Rein:2022odl}
 \\ \hline
 pretzelosity  & $\tilde h_{1T}^\perp(x,b_T)$ & --       & $\cT_h(x)$ & LO $(\as^0)$   & \cite{Gutierrez-Reyes:2018iod,Moos:2020wvd}
 \\ \hline
 \end{tabular}
\caption{Update of collinear matching of the quark TMD PDFs up to collinear twist $3$ at perturbative $b_T^{-1} \gg \LQCD$. 
 The upper four rows of the table show chiral-even TMDs, while the bottom four rows show chiral-odd TMDs.
 $\cT_g(x)$ and $\cT_h(x)$ are abbreviations for specific combinations of the twist-$3$ distributions.}
 \label{tbl:TMDPDF_matching_q}
\end{table}
  \begin{table}
 \centering
 \begin{tabular}{|l|c|c|c|r|c|}
 \hline
 \multirow{2}{*}{Name} & \multirow{2}{*}{Function} & Twist-2  & Twist-3  & \multicolumn{1}{c|}{Known}   & \multirow{2}{*}{Refs.} \\
 & & matching & matching & \multicolumn{1}{c|}{order} &
 \\ \hhline{|=|=|=|=|=|=|}
 unpolarized         & $\tilde f_1(x,b_T)$    & $f_1(x)$ & -- & N$^3$LO $(\as^3)$ &  \cite{Luo:2020epw, Ebert:2020yqt}
 \\ \hline
 linearly polarized  & $\tilde h_1^{\perp g}(x, b_T)$  & $f_1(x)$ & -- & N$^3$LO $(\as^3)$ &  \cite{Zhu:2025ixc}
 \\ \hline
  helicity & $\tilde g_{1L}^g(x, b_T)$ & $g_1(x)$ & & N$^3$LO ($\as^3)$ & \cite{Zhu:2025gts}
 \\ \hline
  & $\tilde g_{1T}^g(x, b_T)$  & & & &
 \\ \hhline{|=|=|=|=|=|=|}
 Sivers & $\tilde f_{1T}^{\perp g}(x, b_T)$  & -- & & &
 \\ \hline
  & $\tilde h_{1T}^g(x, b_T)$  & & & &
 \\ \hline
  & $\tilde h_{1L}^{\perp g}(x, b_T)$  & & & &
 \\ \hline
  & $\tilde h_{1T}^{\perp g}(x, b_T)$  & & & &
 \\ \hline
 \end{tabular}
 \caption{Update of collinear matching of the gluon TMD PDFs up to collinear twist $3$ at perturbative $b_T^{-1} \gg \LQCD$. 
  The upper four rows of the table show chiral-even TMDs, while the bottom four rows show chiral-odd TMDs.
 Empty entries indicate that for most gluon TMDs the matching has not yet been considered in the literature, and these rows presumably obtain their first contributions at twist $3$.
 }
 \label{tbl:TMDPDF_matching_g}
\end{table}

 In addition to the perturbative part of the TMDs, we present new results for the DGLAP splitting functions associated with the three leading-twist operators: the unpolarized, helicity, and quark transversity operators.
 The unpolarized splitting functions are known up to NNLO~\cite{Moch:2004pa,Vogt:2004mw,Mitov:2006ic,Moch:2007tx,Chen:2020uvt} and partially known to N$^3$LO~\cite{Moch:2017uml,Gehrmann:2023ksf,Gehrmann:2023cqm,Gehrmann:2023iah,Falcioni:2023luc,Falcioni:2023vqq,Falcioni:2023tzp,Moch:2023tdj,Basdew-Sharma:2022vya,Falcioni:2024xyt}. Comparing the results with those in the literature~\cite{Mitov:2006ic,Moch:2007tx}, we find full agreement except for the non-diagonal quark-gluon splitting. The discrepancy between our results with those presented in \cite{Almasy:2011eq},
\begin{gather}
\Delta P_{qg}^{{\rm T},(2)}(x) =  P_{qg}^{{\rm T},(2)}\Big|_{\text{this work}} -  P_{qg}^{\text{T},(2)}\Big|_{\text{\cite{Almasy:2011eq}}}  =
\nn
\\
 \frac{\pi^2}{3} (C_F - C_A) \beta_0 
\left[
-4 + 8 x + x^2 + 6 (1 - 2 x + 2 x^2) \ln x 
 \right] \,,
\end{gather}
where $P_{qg}^{\text{T},(2)}$ is the coefficient of $\alpha_s^3/(4 \pi)^3$ in the off-diagonal singlet splitting matrix, and $\beta_0 = 11 C_A/3 - 2 n_f/3$ is the one-loop  QCD beta function.
In Mellin moment space the discrepancy reads
\begin{gather}
   - \int_0^1 dx\, x^{N-1} \Delta P_{qg}^{{\rm T},(2)} (x)
=  (C_A - C_F) \beta_0 \frac{\pi^2}{3} 
\left( \frac{12}{(N+1)^2}
\right.
\nn
\\
 \left. 
 - \frac{6}{N^2} - \frac{12}{(N+2)^2} - \frac{4}{N} + \frac{8}{N+1} + \frac{1}{N+2}  \right) \,.
\end{gather}
The space-like helicity dependent splitting functions were obtained in~\cite{Moch:2014sna,Moch:2015usa,Blumlein:2021enk,Blumlein:2021ryt,Blumlein:2022gpp,Behring:2025avs},
by comparison we find all agreement except for the term with cubic color structure $d_{abc}^2$ for `sea' quark difference $N_f(\Delta P_{qq'}-\Delta P_{q \bar q'})$:
\footnote{Agreement is found at transcendental weight four, whereas discrepancies arise in the weight-three terms.}
\begin{align}
&\Delta P_{d_{abc}^2}^{S,(2)}-\Delta P_{d_{abc}^2}^{\text{Moch:2015}}=
\frac{16 N_f}{3}\frac{d_{abc}^2}{N_c}
\bigg[
-84\zeta_2-18\zeta_3+3H_0+84H_2+36H_3+12H_{-2,0}
\nn\\
+&
(1-z)\times\left(
174 H_1-12H_{1,0,0}
\right)
+
(1+z)\times\left(
36 H_{-1,0}-48 H_{-1,2} -24 H_{-1,0,0}
+48 H_{-1}\zeta_2 -36 H_{0}\zeta_2
\right)
\nn\\
+&
z\times\left(
-174 H_0 +90 H_2 +12 H_3 -24 H_{-2,0} +54 H_{0,0} +24 H_{0,0,0} -54 \zeta_2 -72\zeta_3
\right)
\bigg]\,.
\end{align}
Our results agree with those presented in Ref.~{\cite{Behring:2025avs}}.
Finally, we  verified that our results for NNLO  transversity space-like splitting functions   are consistent with those in literatures~\cite{Vogelsang:1997ak,Mikhailov:2008my,Blumlein:2021enk}.   
\section{Small-$z$ resummation for  unpolarized and linearly polarized gluon TMD FFs}
\begin{figure}[ht!]
\centering
   \includegraphics[width=0.45\textwidth]{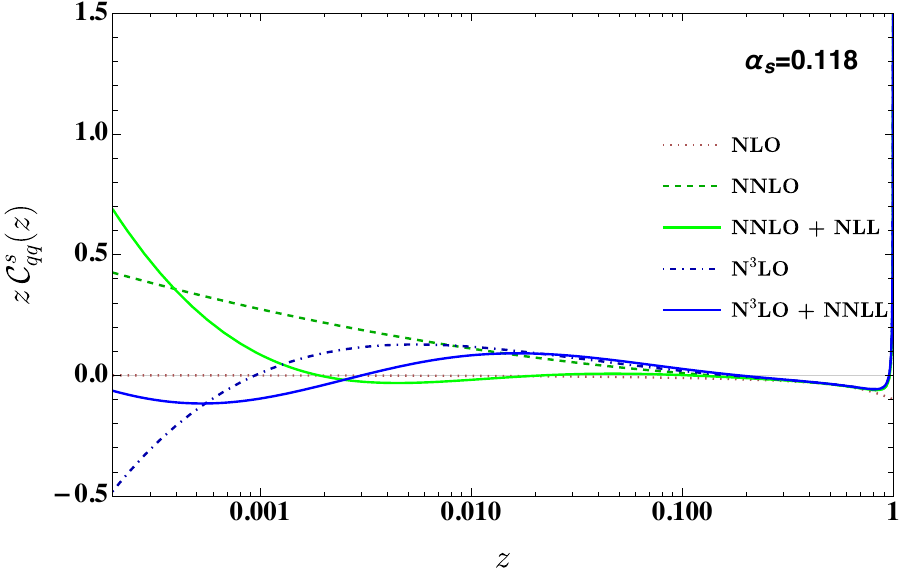}
    \includegraphics[width=0.45\textwidth]{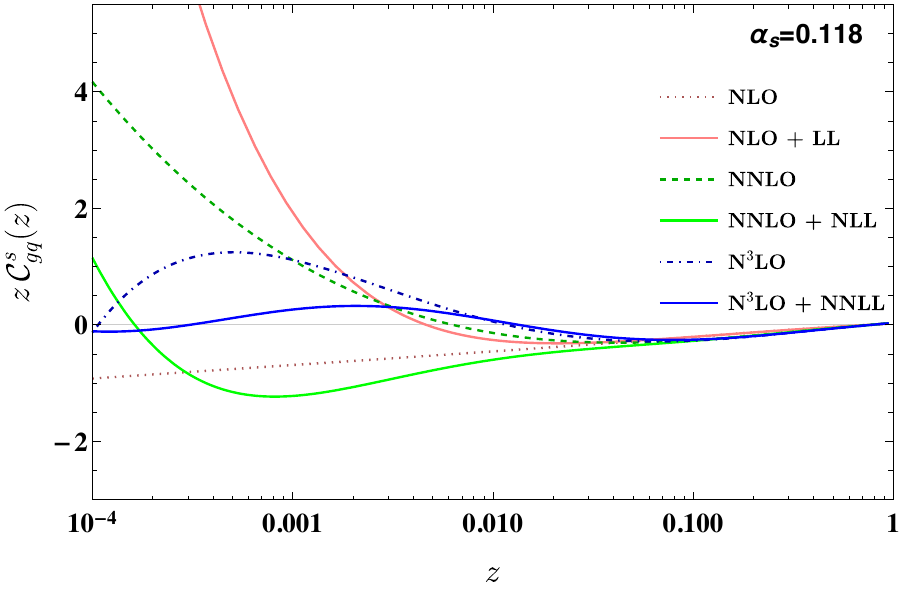}
    \caption{Coefficient functions for quark TMD FFs. Shown in the plots are fixed-order results at NLO, NNLO and  N$^3$LO, as well as adding the higher-order resummation contributions truncated to order $\alpha_s^{15}$.}
\label{fig:Cq2}
 \end{figure}

\begin{figure}[ht!]
\centering
   \includegraphics[width=0.45\textwidth]{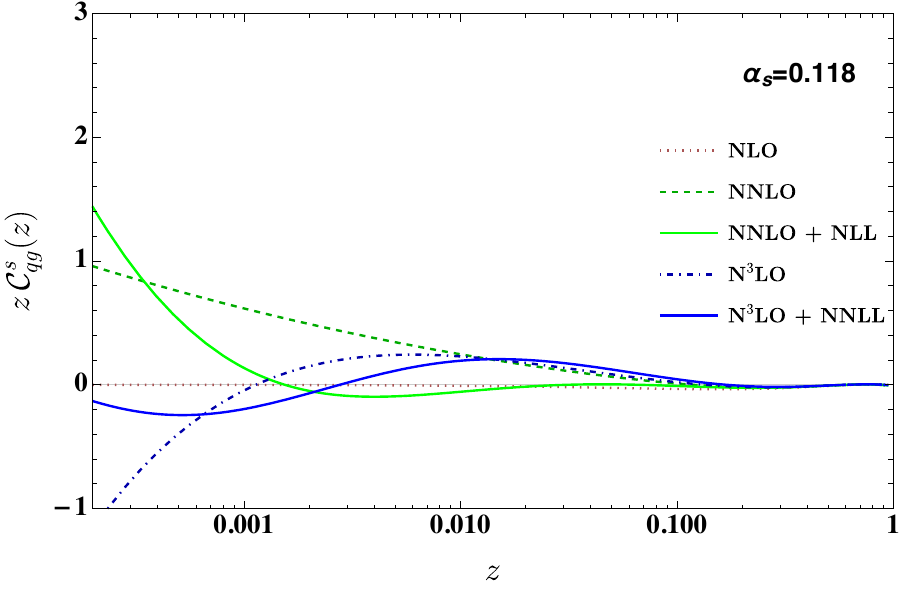}
    \includegraphics[width=0.45\textwidth]{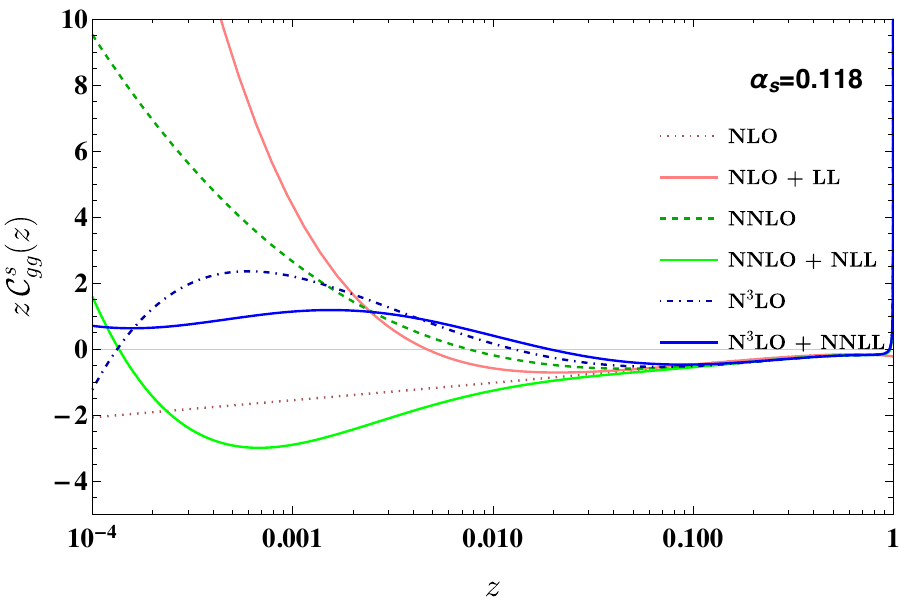}
    \caption{Coefficient functions for gluon TMD FFs. Shown in the plots are fixed-order results at NLO, NNLO and  N$^3$LO, as well as adding the higher-order resummation contributions truncated to order $\alpha_s^{15}$.}
\label{fig:Cg2}
 \end{figure}

\begin{figure}[ht!]
\centering
   \includegraphics[width=0.45\textwidth]{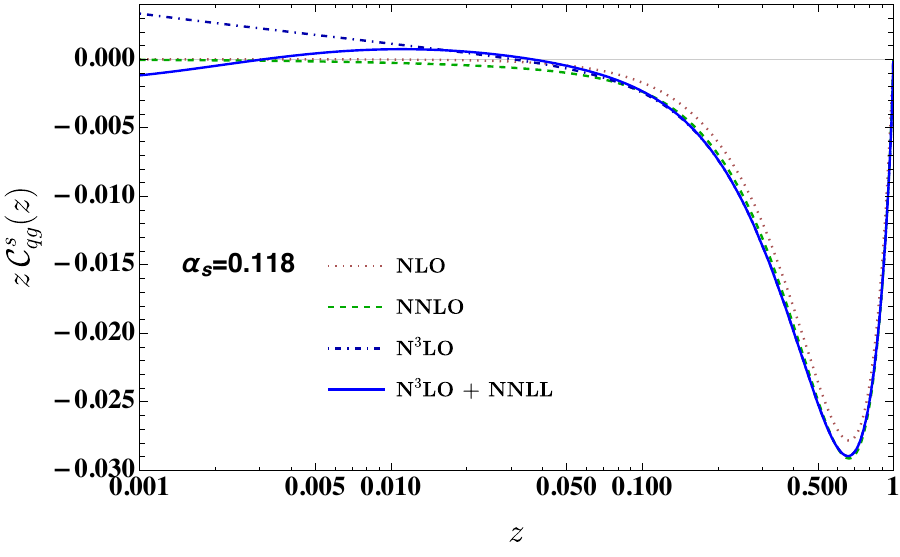}
    \includegraphics[width=0.45\textwidth]{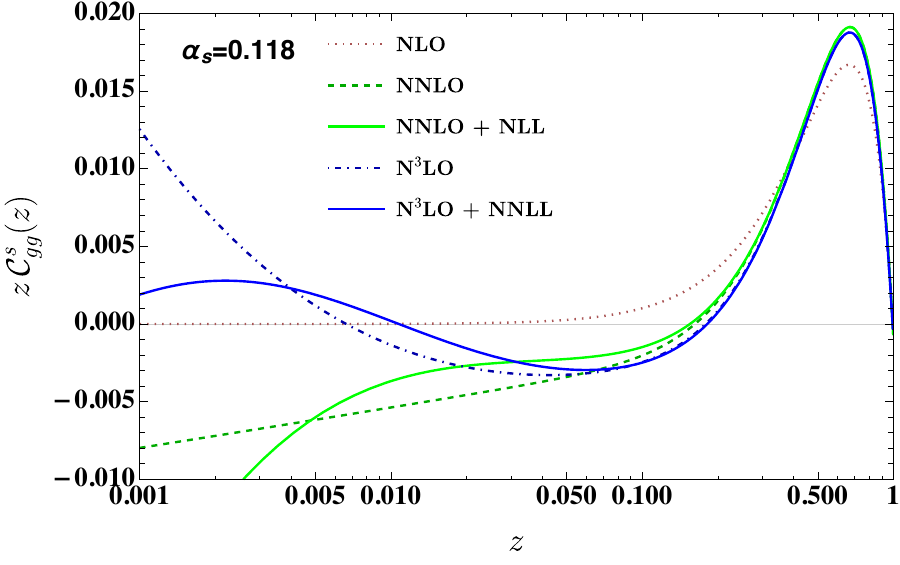}
    \caption{Coefficient functions for the linearly polarized gluon TMD FFs. 
The plots show the fixed-order results at NLO, NNLO and  N$^3$LO, 
together with the resummed predictions including
    higher-order terms truncated at order $\alpha_s^{15}$.}
\label{fig:Cg}
 \end{figure}

Using infrared consistency~\cite{Vogt:2011jv}, we perform small-$z$ resummation for the perturbative parts of the unpolarized and linearly polarized gluon TMD FFs. Figs.~\ref{fig:Cq2} and \ref{fig:Cg2} show the fixed-order coefficient functions for unpolarized TMD FFs with different orders of small-$z$ resummation, while fig.~\ref{fig:Cg} shows the corresponding results for linearly polarized gluon TMD FFs.
 We work in $N_f = 5$ flavor scheme throughout the calculations. We note that even at N$^3$LO, the effect of resummation is important for $z < 10^{-2}$.
\section{Conclusion}
In this work we have completed the N$^3$LO twist-2 matching program for polarized TMD parton distribution and fragmentation functions, including helicity, quark transversity, and linearly polarized gluon TMDs. From these results we extract the complete set of NNLO DGLAP splitting functions in both space-like and time-like kinematics and compare them with existing calculations in the literature. We also investigate the small-$z$ structure of the matching coefficients and perform small-$z$ resummation for  TMD fragmentation functions. 
Together, these developments place polarized TMD observables
on equal theoretical footing with their unpolarized counterparts
and provide key theoretical input for precision spin measurements
at the forthcoming Electron-Ion Collider.
\section*{Acknowledgements}
I thank Hua Xing Zhu for encouraging me to pursue this problem independently and Tong-Zhi Yang for valuable collaboration in the early stages of this project.

{\footnotesize
\bibliography{TMDs}
\bibliographystyle{h-physrev}
}
%

\end{document}